\documentclass[conference,dvipsnames]{IEEEtran}

\usepackage{cite}
\usepackage{amsmath,amssymb,amsfonts}
\usepackage{bbm}
\usepackage{algorithmic}
\usepackage{booktabs}
\usepackage{graphicx}
\usepackage{textcomp}
\usepackage[frozencache,cachedir=minted-cache]{minted}
\usepackage{xcolor}
\usepackage{url}
\usepackage{pifont}
\usepackage{cleveref}
\usepackage{annotate-equations}

\newcommand{\cmark}{\ding{51}}%
\newcommand{\xmark}{\ding{55}}%

\def\BibTeX{{\rm B\kern-.05em{\sc i\kern-.025em b}\kern-.08em
    T\kern-.1667em\lower.7ex\hbox{E}\kern-.125emX}}

\usepackage{bm}
\usepackage{booktabs}
\usepackage{todonotes}
\usepackage{listings}

\begin{document}

\title{Performance-Aligned LLMs for\\Generating Fast Code}

\author{\IEEEauthorblockN{Daniel Nichols\IEEEauthorrefmark{2}, Pranav Polasam\IEEEauthorrefmark{2}, Harshitha Menon\IEEEauthorrefmark{1}, Aniruddha Marathe\IEEEauthorrefmark{1}, Todd Gamblin\IEEEauthorrefmark{3}, Abhinav Bhatele\IEEEauthorrefmark{2}}
\IEEEauthorblockA{~\\
\IEEEauthorrefmark{2}\textit{Department of Computer Science, University of Maryland, College Park, MD, USA}\\
\IEEEauthorrefmark{1}\textit{Center for Applied Scientific Computing, Lawrence Livermore National Laboratory, Livermore, CA, USA}\\
\IEEEauthorrefmark{3}\textit{Livermore Computing, Lawrence Livermore National Laboratory, Livermore, CA, USA}\\
Email: \{dnicho, ppolasam\}@umd.edu, \{marathe1, gopalakrishn1, tgamblin\}@llnl.gov, bhatele@cs.umd.edu}
}

\maketitle

\begin{abstract}
Optimizing scientific software is a difficult task because codebases are often
large and complex, and performance can depend upon several factors including the
algorithm, its implementation, and hardware among others. Causes of poor
performance can originate from disparate sources and be difficult to diagnose.
Recent years have seen a multitude of work that use large language models (LLMs)
to assist in software development tasks. However, these tools are trained to
model the distribution of code as text, and are not specifically designed to
understand performance aspects of code. In this work, we introduce a
reinforcement learning based methodology to align the outputs of code LLMs with
performance. This allows us to build upon the current code modeling capabilities
of LLMs and extend them to generate better performing code. We demonstrate that
our fine-tuned model improves the expected speedup of generated code over base
models for a set of benchmark tasks from 0.9 to 1.6 for serial code and 1.9 to
4.5 for OpenMP code.

\end{abstract}


\section{Introduction}
\label{sec:introduction}
Developing fast and scalable code is a difficult, but often necessary task for
scientific software developers. It can require expert knowledge of the
application domain, algorithm design, programming languages, and hardware. This
is a challenging task for even serial code, and even more complex for parallel
code.  Further, programmers and performance engineers are often tasked with
optimizing existing code, often not written by them, which requires
understanding an existing codebase and the performance implications of changes.
Large language models (LLMs) have emerged as a powerful tool for assisting in
the software development process for a variety of tasks such as code
completion~\cite{codex-copilot-short-author}, bug
detection~\cite{Richter2022CanWL, Kharkar2022LearningTR}, and code
summarization~\cite{Ahmad2020ATA, Haque2022SemanticSM, Gu2022AssembleFM,
Ahmed2022LearningCS}. Recently, they have also been used with limited success to
generate parallel code~\cite{nichols:arxiv2023}. Yet they struggle to understand
performance aspects of code because they were not designed for this task.  Code
LLMs are trained on just code as text, and as a result, are not well-suited to
reason about complex performance issues.  Additionally, the code they generate
does not consider performance and could be slow, despite being correct. This has
been demonstrated in existing works that show LLMs often generate inefficient
parallel code~\cite{nichols:arxiv2024, valerolara2023comparing}. 

Creating artificial intelligence (AI) models that can generate faster code has
the potential to significantly improve the productivity of software developers.
By using performance-aware code LLMs, developers can focus on design and
correctness without worrying about the performance implications of using LLMs to
generate code. Additionally, as LLM-based tools become more integrated with
software development workflows, developers will become more and more reliant on
the quality of their outputs. Improving the performance of LLM generated code
while maintaining its correctness will improve the quality of the target
software being developed. Further, code LLMs that can write fast code can remove
the need for every scientific and parallel programmer to be a performance expert
in addition to their existing domain expertise.

It is non-trivial to create code LLMs that can generate faster code. Since
creating performance-aware code LLMs will require fine-tuning of LLMs using
performance data, one challenge is creating such datasets. LLMs typically
require very large, general datasets for training tasks, and it is challenging
to create such large datasets for performance data. Arbitrary code can have a
wide range of performance characteristics, and depend on many factors such as
input data, hardware, and software environment. Due to the complexity in
collecting performance data for arbitrary code, performance datasets are often
small and/or narrow in focus. Further, even with such a dataset in hand, an LLM
needs to be carefully fine-tuned to {\em align} its generated outputs with more
performant code. There are many potential pitfalls here, for instance, improving
the performance of generated code at the cost of correctness. Additionally,
fine-tuned LLMs can learn a distribution too disjoint from their initial code
distribution they modeled and lose their ability to generalize.

In order to overcome the challenges associated with collecting large scale
performance data, we propose a new approach that combines a structured, narrow
performance dataset with a more general synthetic code dataset for fine-tuning.
We also propose two novel fine-tuning methodologies:~(1) reinforcement learning
with performance feedback (RLPF), which is based on reinforcement learning with
human feedback (RLHF)~\cite{ouyang2022training}, and direct performance
alignment (DPA), which is based on direct performance optimization
(DPO)~\cite{rafailov2023direct}. We use these two approaches and the new dataset
to align an existing code LLM to generate faster code. These proposed
fine-tuning methodologies use fast and slow code pairs to fine-tune the LLMs to
generate samples more similar to the fast code and less similar to the slow
code. The aligned model is then evaluated on two code generation benchmarks and
one code optimization benchmark. We find that the aligned model is able to
generate code with higher expected speedups than that of the original model,
while maintaining correctness.

This work makes the following important contributions:
\begin{itemize}
	\item A code performance dataset that combines narrow, structured
		performance data with broad synthetic data to help models
		learn performance properties, but maintain their ability
		to generalize.
	\item Two novel fine-tuning methodologies, reinforcement learning with
		performance feedback (RLPF) and direct performance alignment (DPA),
		for aligning code LLMs to generate faster code.
    \item A fine-tuned, performance-aligned LLM that generates faster code than
		traditional code LLMs.
    \item A detailed study of the performance and correctness of the code
		generated by performance-aligned LLMs including serial, OpenMP, and MPI
		code. Additionally, an ablation study motivating the use of synthetic
		data to fine-tune code LLMs for performance.
\end{itemize}

\section{Background}
\label{sec:background}
In this section, we provide a background into large language models and their
use for code generation. We further provide an overview of reinforcement
learning and the Proximal Policy Optimization algorithm.

\subsection{Large Language Models for Code}\label{sec:bf-llms-for-code}

LLMs have been shown to be effective tools for many code generation
tasks~\cite{codex-copilot-short-author, li2023starcoder, li2022competition}.
These LLMs are typically Transformer models~\cite{transformer} fine-tuned on large
code datasets~\cite{guo2024deepseekcoder, wei2023magicoder, li2023starcoder} to
model the probability distribution of code text data. These models can then be
used to generate code, fill in missing code snippets, complete code snippets,
and more. Code is generated by showing them a sequence of code text (as tokens)
and using the model to predict the next token in the sequence. Getting good text
generation with this method is not always straightforward, so additional
sampling techniques such as {\it temperature} and {\it top-p} are often used to
improve the quality of the generated text~\cite{holtzman:iclr2020}. These
control the randomness of the sampling process, with {\it temperature}
controlling the entropy of the distribution and {\it top-p} controlling the
number of tokens considered for sampling.

\subsection{Reinforcement Learning and Proximal Policy Optimization}\label{sec:bf-rl-and-ppo}

Reinforcement learning (RL) is a popular machine learning training paradigm
where an agent model learns to interact with an environment to maximize a reward
signal. This learning is typically accomplished by the agent iteratively taking
actions in the environment, observing the results, and updating its policy to
maximize the reward. While RL techniques have been popular for a number of
years, they have recently been applied to LLMs due to their success in
aligning LLM outputs with human preferences~\cite{ouyang2022training}.

Proximal Policy Optimization (PPO)~\cite{schulman2017proximal} is a popular RL
algorithm that has been used to successfully fine-tune LLMs. It is a
state-of-the-art algorithm that has become widely used due to its efficiency and
robustness across a number of different tasks. A key difference between PPO and
other RL algorithms is its use of clipping to prevent unusually large updates to
the policy. PPO clips the ratio of the new agent policy and the previous agent
policy to a range of $[1-\epsilon, 1+\epsilon]$. This prevents large weight
updates, which can lead to instability in the training process. The clipped
policy updates are combined with a value loss function (the reward signal) and
an entropy loss function (to encourage exploration) to train the agent. After
running many iterations of the training process, the agent learns to make
decisions that optimize the reward signal. In this paper, we will train an agent
(an LLM) to generate code that is fast (higher reward for faster, correct code).

\section{Overview of Methodology}
\label{sec:overview}
\begin{figure}[h]
    \centering
    \includegraphics[width=\columnwidth]{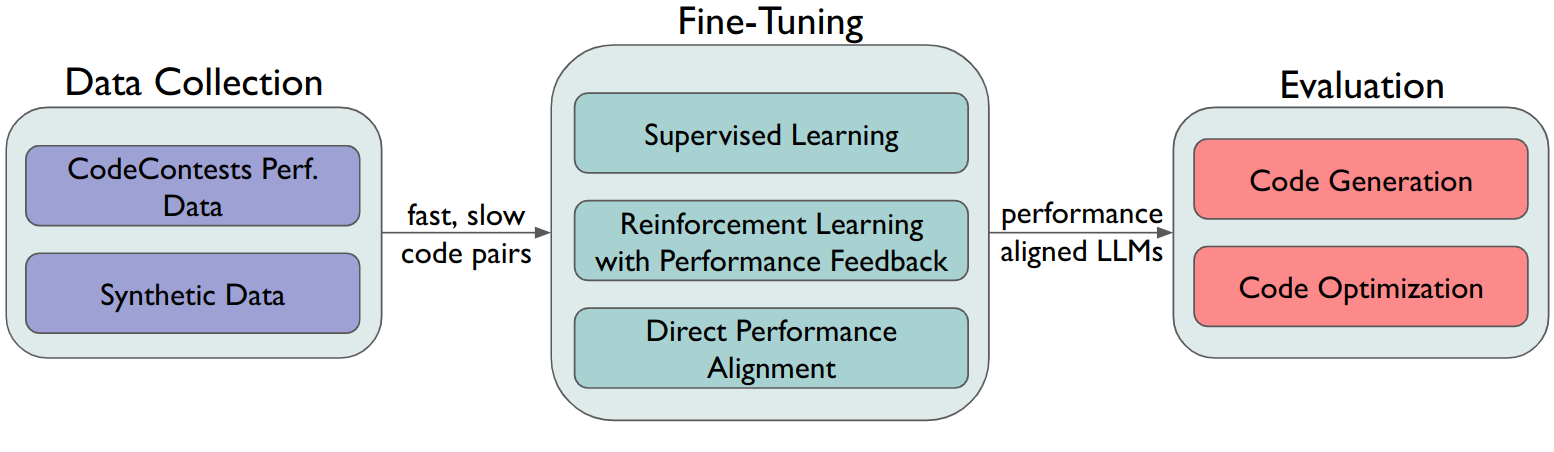}
    \caption{An overview of the proposed methodology. We first collect a large
    dataset of fast and slow code pairs using coding contest submissions and 
    synthetically generated data. Then we fine-tune three different LLMs 
    on this data to generate faster code. Finally, we evaluate the fine-tuned
    models on code generation and optimization tasks.\label{fig:overview}}
\end{figure}

\Cref{fig:overview} presents an overview of our methodology for aligning code
large language models (LLMs) to generate faster code. We start by creating a
dataset that can be used to fine-tune an LLM to generate code that is both
correct and fast (\Cref{sec:data}). To accomplish this, we collect a large,
structured code dataset with performance data and test cases to measure
correctness. This structured dataset is, however, not representative of the
entire distribution of code we want an LLM to optimize so we ameliorate its
shortcomings by using LLMs to generate a {\em synthetic} code dataset that
covers a wider range of code.

These datasets are then used to {\em align} the outputs of an LLM with
performance considerations. We employ three different techniques -- supervised
learning, reinforcement learning, and direct alignment, to fine-tune code LLMs
(\Cref{sec:rl}). The models are aligned to answers that are not only correct,
but also fast. Using the fine-tuned models we then generate code for a set of
three different benchmark tasks for code generation and optimization
(\Cref{sec:eval}). These tasks measure the correctness and performance of the
generated code for coding problems within and outside the distribution of the
training data.

\section{Data Collection and Labeling}
\label{sec:data}
In order to align LLMs to generate more performant output, we need to fine-tune
them on performance data. Further, to apply the proposed fine-tuning methods, we
need a dataset of code where we have a slow and a fast implementation of a
particular problem. This type of structured performance data paired with source
code is difficult to collect. It requires being able to build, execute,
validate, and profile arbitrary code snippets, which is difficult to accomplish
at scale. In this section, we describe our process of collecting a large
performance dataset ($\mathcal{D}_c$). Additionally, we discuss how we extend
the dataset with synthetic data ($\mathcal{D}_s$) to cover a wider distribution
of code patterns. The final dataset $\mathcal{D}$ contains over 4.5 million code
samples, distributed over three source languages (C++, Java, and Python) as
shown in Table~\ref{tab:dataset-langs}.

\begin{table}[h]
    \centering
    \caption{The number of samples in both datasets distributed by source language.}
    \label{tab:dataset-langs}
    \begin{tabular}{@{}lccccc@{}}
    \toprule
    \textit{\textbf{Dataset}} ($\mathcal{D}$)             & \begin{tabular}[c]{@{}c@{}}Runtime\\ Data\end{tabular} & C++       & Java    & Python    & \begin{tabular}[c]{@{}c@{}}No. of\\ Samples\end{tabular}     \\ \midrule
    \textit{CodeContests+Perf} ($\mathcal{D}_{c}$)        & \cmark                                                     & 1.8M & 0.9M & 1.8M & \textbf{4.5M} \\
    \textit{Synthetic} ($\mathcal{D}_{s}$)                & \xmark                                                     & 5k      & 0       & 5k      & \textbf{10k}    \\ \bottomrule
\end{tabular}
\end{table}

\subsection{Performance Dataset Collection}\label{sec:perf-dataset}

We build our performance dataset using the CodeContests dataset introduced by
DeepMind in~\cite{li2022competition}. This dataset contains coding contest
problems and solutions from the Aizu~\cite{aizu}, AtCoder~\cite{atcoder},
CodeChef~\cite{codechef}, Codeforces~\cite{codeforces}, and
HackerEarth~\cite{hackerearth} online competition platforms. In total there are
13,610 coding problems in the dataset. These range in difficulty from simple to
very difficult, and cover a wide range of topics such as graph algorithms,
dynamic programming, and search. Each problem in the dataset has a
corresponding set of submissions from users, labeled as correct or incorrect on
the respective coding contest website. The number of submissions per problem
ranges between tens and thousands. There are solutions in three different
programming languages: C++, Java, and Python. Additionally, the dataset
includes meta-data for the problem such as the problem statement, test cases,
time limits, and memory limits.

This dataset is extremely valuable for our study as it provides a large amount
of code samples along with the necessary tests to measure correctness and
performance. More so, it contains many code samples that solve the same problem,
but in different ways and with different runtimes. While many of the code
contest websites record runtimes for submissions, the CodeContests dataset as
provided by DeepMind does not include this information. We collect this data
ourselves into a new dataset, {\it CodeContests-Perf} ($\mathcal{D}_c$), by
executing each of the correct submissions and recording their runtimes. Each
submission is run on all the test cases for its problem. Generally, there are
between 5 and 20 test cases per problem. We create submission-runtime pairs
using the average runtime over all the test cases.  Each run is executed on a
single core of an AMD EPYC 7763 CPU with a 2.45 GHz base frequency.

The final {\it CodeContests-Perf} dataset contains 4.5 million samples. The
distribution of samples by source language is shown in
Table~\ref{tab:dataset-langs}. There were a small fraction of submissions
labeled as correct in the CodeContests dataset that errored or failed the test
cases when we ran them. These are omitted from the final dataset. We also
include code submissions that were marked as incorrect in the original dataset,
however, we do not run them. These will eventually be useful to prevent the
model from generating fast, but incorrect code.

\subsection{Synthetic Data Generation}\label{sec:synthetic-data}

The amount of data and the availability of easy testing in the {\it
CodeContests-Perf} dataset makes it a crucial component of our study. However,
the distribution of code represented in the dataset is significantly different
than that of the code that is typically found in production code. Coding
contests generally award participants based on time-to-submission leading to
users writing messy and/or disorganized code to solve problems as quickly as
possible. Further, the types of problems typically found in coding contests such
as depth-first search and dynamic programming, while an important subset of
problems, do not cover the full range of relevant computational problems that
are found in production code, and in particular, in scientific computing.

To address the shortcomings of the {\it CodeContests-Perf} data, we generate an
additional synthetic dataset $\mathcal{D}_s$ of fast and slow code samples. This
is inspired by several recent works demonstrating the effectiveness of
fine-tuning LLMs on synthetic data to improve performance on real
tasks~\cite{wei2023magicoder, zheng2023judging, Gilardi_2023, he2023annollm,
benallal2024cosmopedia}. Gilardi et al.~\cite{Gilardi_2023} even find that LLMs
can outperform humans for many text annotation tasks. In our case of annotating
code performance, real runtimes are the best annotation, but in the absence of
runtime data, synthetic data is a promising candidate to obtaining labeled code
performance data.

We use the Gemini-Pro-1.0 LLM model~\cite{geminishort2023gemini} to generate
synthetic code samples as we found it to give the best outputs among a number of
models we tested. We adapt the methodology in~\cite{wei2023magicoder}, where
samples are generated using {\it seed} code snippets to get diverse outputs from
the model. First, we create a dataset of 10,000 seed samples that are 1-15 line
random substrings of random files from The Stack
dataset~\cite{Kocetkov2022TheStack}, which is a large, 3TB dataset of
permissively licensed code. Then the LLM is asked to generate three pieces of
text: a problem statement inspired by the seed snippet, a fast solution to the
problem, and a slow solution to the problem. This produces inherently noisy
data, since the LLM does not always generate correct or optimal (fast vs.~slow)
outputs. However, prior work has shown that the gain in predictive performance
from fine-tuning on synthetic data often outweighs the downsides from noisy
data~\cite{wei2023magicoder}.

In total, we collect 10,000 synthetic samples, 5,000 in C++ and 5,000 in Python.
While adding more synthetic samples would likely continue to improve the quality
of the fine-tuned model, we found that limiting to 10,000 samples provided
adequate model quality while operating within time/cost constraints for this
study. Table~\ref{tab:dataset-langs} shows the distribution of samples by
language in the synthetic dataset.

\section{Aligning LLMs to Generate Faster Code: Proposed Fine-Tuning Approaches}
\label{sec:rl}
Large language models have been shown to be capable of generating correct code
with high frequency on several benchmarks~\cite{codex-copilot-short-author,
mbpp, openai2023gpt4}, yet they do not always generate code that is
efficient~\cite{nichols:arxiv2024}. They require further fine-tuning to align
them with performance considerations. In this section, we detail how we
fine-tune large language models with supervised learning and reinforcement learning techniques to generate faster
code. We utilize the dataset introduced in Section~\ref{sec:data} to train
three different models using supervised learning, reinforcement learning with
performance feedback, and direct performance alignment.

\subsection{Supervised Learning}\label{sec:supervised}

In the first approach, we fine-tune a language model on the dataset of code
snippets from $\mathcal{D}$ to predict the next token in a sequence given
previous tokens. For our methodology, we begin with a model that has already
been trained on a large corpus of text and code, and then fine-tune it on a
smaller dataset of coding problems and fast solutions.

We create two types of prompts using the samples in $\mathcal{D}$ to fine-tune
the model. In the first type of prompt, we use a standard instruction prompt
where the model is given a problem statement and a fast solution (shown
in~\Cref{lst:sft-prompt}).  Using the coding contest data in $\mathcal{D}_c$,
we use the problem description as the instruction and randomly sample one of
the five fastest solutions as the response.  In the second type of prompt, we
use a variation of the standard instruction prompt where the task is to
optimize a given code snippet and the output is an optimized version of the
code. For this, we use the problem description and one of the slowest 33\% of
solutions as the instruction, and one of the five fastest solutions as the
response.  Forming prompts from the synthetic dataset $\mathcal{D}_s$ is
similar except we only have one slow and one fast solution for each problem, so
we do not sample from ranges of solutions.

\definecolor{LightGray}{gray}{0.9}
\definecolor{LightBlue}{RGB}{240, 248, 255}
\begin{listing}[ht]
\begin{minted}[fontsize=\small,bgcolor=LightBlue]{markdown}
### Instruction:
Given a list of strings, find the longest
common prefix shared by all strings in the 
list. The prefix should be the longest 
possible string that is a prefix of every 
string in the list.

### Response:
```python
def longest_common_prefix(strings):
  if not strings:
    return ""
    
  prefix = strings[0]
  for string in strings[1:]:
    while string.startswith(prefix):
      prefix = prefix[:-1]
  return prefix
```
\end{minted}
\caption{The instruction prompt format used to fine-tune the models. During
fine-tuning, a coding problem is given to the model as an instruction-response
pair, and the model is trained to generate similar responses when used for
inference.}
\label{lst:sft-prompt}
\end{listing}

Over these prompts, the model is fine-tuned to minimize the cross-entropy loss
between its predicted next token and the actual next token. We refer the reader
to~\cite{naveed2024comprehensive} for more details on fine-tuning LLMs for text
generation. After fine-tuning, the model should have more fast code snippets in
its training data and its probability distribution should shift toward
faster code. Several prior works, however, have observed that methods more
sophisticated than supervised fine-tuning are required to align LLM outputs with
certain properties, such as safety and human
preferences~\cite{ouyang2022training, ziegler2020finetuning, bai2022training}.

\vspace{0.07in}
\noindent{\bf Supervised Fine-Tuning Evaluation Metric}:
We evaluate the success of the supervised fine-tuning by measuring the
perplexity of the tuned model over an evaluation dataset. Perplexity is
inversely proportional to how confident a model is that a data sample is in the
distribution it models. A lower perplexity is better and indicates the LLM is
less ``perplexed'' by a particular sample. A model's perplexity over $t$ tokens
from a dataset $\mathcal{X}$ is given by~\Cref{eq:perplexity}.

\vspace{1em}
\begin{equation}
    \label{eq:perplexity}
    \text{Perplexity}(\mathcal{X}) =
    \exp \left\{
        -\frac1t
        \sum_{i}^{t} 
        \log 
        \eqnmarkbox[MidnightBlue]{pt}{p_{\theta} \left( x_i \mid x_{<i} \right)}
    \right\}
\end{equation}
\annotate[yshift=1.2em]{above,left}{pt}{Predicted probability of token $x_i$ \\ given the previous tokens $x_{<i}$}

\subsection{Reinforcement Learning with Performance Feedback}\label{sec:rlpf}

To further align an LLM's outputs with performance considerations, we propose a
new method, which we call reinforcement learning with performance feedback.
This method is inspired by the success of reinforcement learning with human
feedback (RLHF)~\cite{ouyang2022training}, which aligns LLM outputs with human
preferences. RLHF uses human-labeled preference data to train a reward model
that assigns rewards to LLM outputs that are more preferred by humans. This
reward model is used in conjunction with reinforcement learning to fine-tune a
LLM to generate outputs that are more preferred by humans. We adapt this method
into reinforcement learning with performance feedback (RLPF) that uses
performance feedback instead of human feedback to fine-tune LLMs to generate
faster code.

\vspace{0.07in}
\noindent{\bf Reward Model}:
We first need to design a reward function that can be used to guide the
reinforcement learning process. If we can automatically run, test, and
measure the performance of a generated LLM output, then we can simply use a
function of the recorded runtime as the reward. In our case, this is possible
for the coding contests dataset $\mathcal{D}_c$, where we have unit tests
available to run and test the generated code (see~\Cref{sec:perf-dataset}).
This further highlights the utility of this dataset for our study. 

As mentioned in~\Cref{sec:synthetic-data}, we want to be capable of generating
fast code outside the context of coding contests~i.e.~we do not want to
exclusively use the code contests data for RL fine-tuning. Since we may not be
able to obtain runtime data for other arbitrary code samples, we need to train a
reward model that rewards faster code more than slower code for samples where we
cannot obtain runtime data. Fine-tuning LLMs for relative performance modeling
was previously demonstrated by Nichols et al.~\cite{nichols:arxiv2023} and, thus,
a fine-tuned LLM is a viable candidate for the reward model.

To accomplish this we train a reward model (an LLM), $r_\theta$, to predict a
reward for a given code sample, where a higher reward indicates faster code. To
train this model, we first use a subset of $\mathcal{D}$ to create a dataset of
triplets $(p, d_f, d_s)$ where $p$ is a problem description and $d_f$ and $d_s$
are fast and slow code solutions to the problem, respectively. Using $r_\theta$,
we compute predicted rewards for $d_f$ and $d_s$, and use these to
calculate the loss function $\mathcal{L}_r$ in~\Cref{eq:reward-loss}.
 
\vspace{1.75em}
\begin{equation}\label{eq:reward-loss}
    \mathcal{L}_{r} = - \log 
    \left[
        \sigma\left(
            \eqnmarkbox[OliveGreen]{df}{r_\theta\left( p, d_f \right)} -
            \eqnmarkbox[ProcessBlue]{ds}{r_\theta\left( p, d_s \right)} -
            \eqnmarkbox[Plum]{mu}{\mu\left( p, d_f, d_s \right)}
        \right)
    \right]
\end{equation}
\annotate[yshift=1em,xshift=-0.2em]{above,left}{df}{predicted reward\\for fast code}
\annotate[yshift=1em,xshift=0em]{above,right}{ds}{predicted reward\\for slow code}
\annotate[yshift=-0.5em]{below,left}{mu}{adaptive margin;\\scales the reward based on runtimes}
\vspace{0.8em}

This loss function is used to train $r_\theta$ to predict a higher reward for
$d_f$ than $d_s$.  In~\Cref{eq:reward-loss} $\sigma$, is the logistic function
and $\mu$ is an adaptive margin as defined in~\Cref{eq:reward-loss-margin}.
The loss function in~\Cref{eq:reward-loss} is adapted from Wang et
al.~\cite{wang2024secrets} to include runtime information. It trains the reward
model to generate rewards farther and farther apart for faster and slower code
samples. As $r_\theta(p,d_f) - r_\theta(p,d_s)$ gets larger, the loss function
tends towards zero. On the flip side, the loss increases as the difference
between the rewards decreases or $r_\theta$ assigns a larger reward to the
slower code. We utilize an adaptive margin $\mu$ to further scale
the rewards based on how much faster the fast code is than the slow code:

\vspace{1em}
\begin{equation}\label{eq:reward-loss-margin}
    \mu\left( p, d_f, d_s \right) = 
    \begin{cases}
        \min\left\{
            \eqnmarkbox[SeaGreen]{lambda}{\lambda}, 
            \eqnmarkbox[OrangeRed]{speedup}{\frac{
                \mathrm{runtime}(d_s)
            }{
                \mathrm{runtime}(d_f)
            }}
        \right\} & \textrm{if } p \in \mathcal{D}_c \\
        0  & \text{otherwise} \\
    \end{cases}    
\end{equation}
\annotate[yshift=1.0em]{above,left}{lambda}{max margin value}
\annotate[yshift=1.0em]{above,right}{speedup}{speedup of $d_f$ over $d_s$}

\begin{figure}[ht]
    \centering
    \includegraphics[width=0.9\columnwidth]{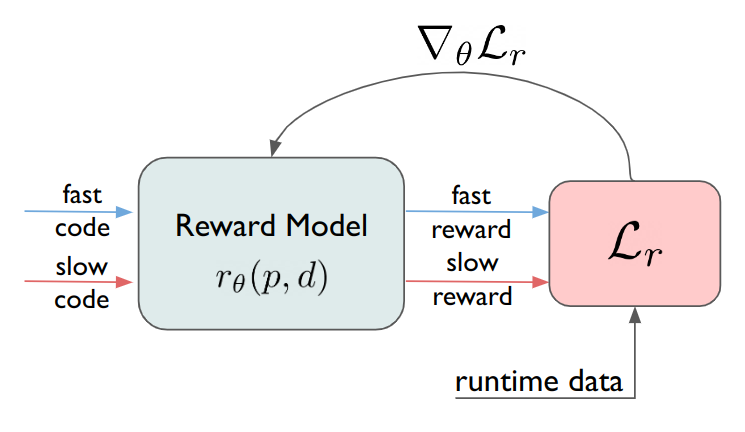}
    \caption{\label{fig:reward-fine-tuning}
    An overview of the reward model fine-tuning process. The reward model 
    outputs a reward for a fast and slow code sample. The loss function uses
    these rewards alongside runtime data to update the weights of the model so
    that its predicted rewards move farther apart for faster and slower code
    scaled by the runtime speedup.}
\end{figure}

Since we can train the reward model on both datasets $\mathcal{D}_c$ and
$\mathcal{D}_s$, we can use the runtime information from $\mathcal{D}_c$ to
scale the rewards appropriately. We use a max margin $\lambda$ to prevent
extremely large margins when $d_s$ is very slow. \Cref{fig:reward-fine-tuning}
provides an overview of the reward model fine-tuning process.

It is important to note that the reward model $r_\theta$ is not directly
modeling code performance. Doing so would likely be impossible as performance
can depend on a number of factors like hardware, input, etc. that are not
accounted for in the input to the reward model. Instead, the reward model is
trained to learn code structures and patterns that generally lead to better
performance. This is another reason it is important to have a large dataset
that covers a wide distribution of code, so that the model can learn these
generalizations.

Using the runtime data in $\mathcal{D}_c$ and the trained reward model, we can
define a reward function $r(p,d)$ that assigns a reward to an LLM generated
code sample. This reward function is defined in~\Cref{eq:reward}.

\begin{equation}\label{eq:reward}
    r(p, d) =
    \begin{cases}
        -1 & \textrm{if } p \in \mathcal{D}_c,\, d \textrm{ incorrect} \\
        \frac{\mathrm{median\_runtime}(p)}{\mathrm{runtime}(d)}-1 & \textrm{if } p \in \mathcal{D}_c,\, d \textrm{ correct} \\
        r_\theta(p, d) & \text{otherwise} \\
    \end{cases}
\end{equation}

The model is penalized with a negative reward if it generates incorrect code.
If it generates correct code, then the reward is based on the speedup over the
median runtime, $\mathrm{median\_runtime}(d)$, from the submission already in
the dataset.  For the synthetic problems, we use the output of the reward model
$r_\theta$.

\vspace{0.07in}
\noindent{\bf Reward Model Fine-Tuning Evaluation Metric}:
We can evaluate the fine-tuning of the reward model by computing its accuracy
over an evaluation dataset. The accuracy here is defined as the proportion of
samples where the reward signal is larger for the fast code than it is for the
slow code.

\begin{equation}\label{eq:reward-accuracy}
    \text{acc}_{\mathrm{reward}}(\mathcal{X}) = \frac{1}{|\mathcal{X}|} \sum_{(p, d_f, d_s) \in \mathcal{X}} 
    \mathbbm{1}\left[ r_\theta(p, d_f) > r_\theta(p, d_s) \right]
\end{equation}

Here $\mathbbm{1}$ is the indicator function that returns 1 if the condition is
true and 0 otherwise. A perfect accuracy of 1 indicates that the reward model
always predicts a higher reward signal for the fast code sample than the slow
code sample.

\vspace{0.15in}
\noindent{\bf Reinforcement Learning}:
Using the reward function $r(p, d)$ and Proximal Policy Optimization
(PPO)~\cite{schulman2017proximal}, we can align an LLM to generate faster code.
We use the supervised fine-tuned model from~\Cref{sec:supervised} as the base
model to fine-tune with RL as is common in RLHF~\cite{ouyang2022training}.
Following standard PPO training practices we optimize the base model using the
reward objective function in~\Cref{eq:ppo-reward-objective}.

\vspace{2em}
\begin{equation}\label{eq:ppo-reward-objective}
    \mathcal{L}_p = r(p,d) - \eta \mathrm{KL}\left(
        \eqnmarkbox[WildStrawberry]{rlpf}{\pi^{\mathrm{RLPF}}(d \mid p)}
        \parallel 
        \eqnmarkbox[Emerald]{sft}{\pi^{\mathrm{S}}(d \mid p)}
    \right)
\end{equation}
\annotate[yshift=0.75em]{above,left}{rlpf}{new model $\pi^{\mathrm{RLPF}}$\\being fine-tuned with RL}
\annotate[yshift=0.75em,xshift=-2.5em]{above,right}{sft}{supervised model $\pi^\mathrm{S}$}

Here KL is the Kullback-Leibler divergence and $\eta$ is a hyperparameter that
controls the divergence penalty. This penalty helps prevent the model from
getting stuck in local optima or diverging too far from the original
distribution of the supervised model~\cite{jaques2019way,
laidlaw2023preventing}.

\begin{figure}[h]
    \vspace{-1em}
    \centering
    \includegraphics[width=\columnwidth]{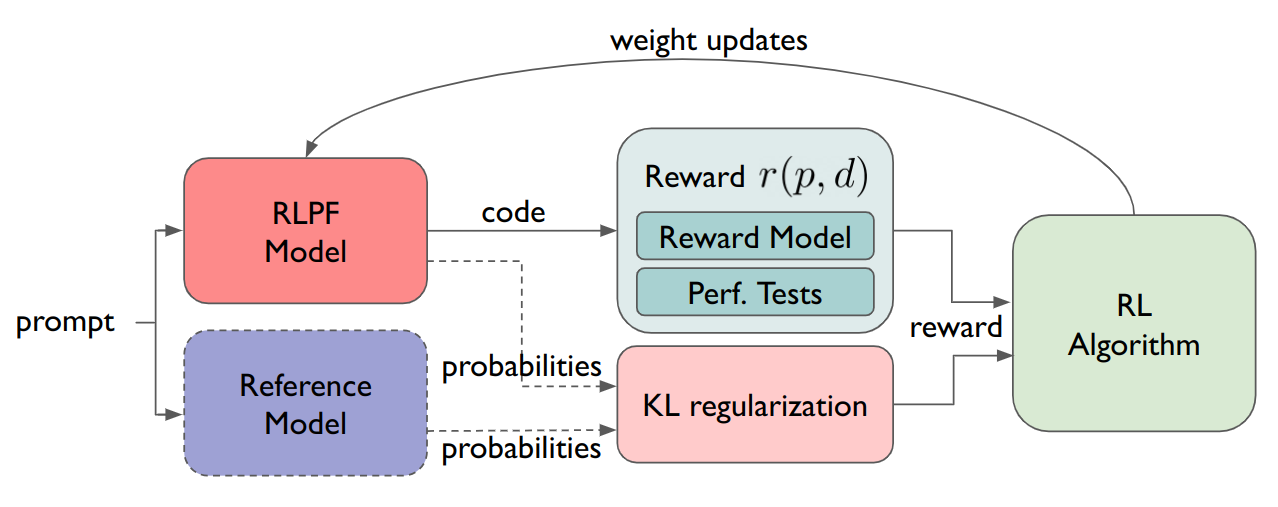}
    \caption{\label{fig:rlpf-fine-tuning}
    The RLPF fine-tuning process. A prompt is given to the model and a reward is
    calculated based on the code it generates. Additionally, the KL-divergence
    between a reference model and the fine-tuned model is included in the reward
    to prevent deviating too far from the original distribution. Finally, PPO is
    used to update the model's parameters based on the reward.}
\end{figure}

During fine-tuning, a prompt is given to the base model (a coding problem or
optimization task) and is used to generate a response. The reward function $r(p,
d)$ is then used to compute a reward for the response either by running the
generated code or getting a reward from the reward model. The reward is then
used to compute the loss function $\mathcal{L}_p$
in~\Cref{eq:ppo-reward-objective}. The loss is then used to update the base
model's parameters using PPO. The process is repeated for a number of iterations
$T$ or until the model converges.  \Cref{fig:rlpf-fine-tuning} provides an
overview of the RLPF fine-tuning process.

\vspace{0.07in}
\noindent{\bf RLPF Fine-Tuning Evaluation Metric}: We can measure the success of
the reinforcement learning using two metrics: the mean reward and the magnitude
of the KL-divergence over an evaluation dataset.  The mean reward indicates
how well the LLM being fine-tuned is able to optimize the reward function. A
higher mean reward is better and indicates that the model is generating faster
code. The KL-divergence measures how far the fine-tuned model has diverged from
the supervised model. The absolute magnitude of this is difficult to interpret,
but it should remain positive and low to indicate that the fine-tuned model is
not diverging too far from the supervised model.

\subsection{Direct Performance Alignment}\label{sec:dpa}

In recent work, Rafailov et al.~\cite{rafailov2023direct} demonstrated an
alternative approach that does not use reinforcement learning to align LLM
outputs with certain properties. Their approach, called Direct Preference
Optimization (DPO), uses a derivation of RLHF's reward objective (similar
to~\Cref{eq:ppo-reward-objective}) to directly update the model's parameters to
align with a reward signal, rather than train a reward model and use RL. The
derived loss takes a similar form to the reward loss in~\Cref{eq:reward-loss}.
This DPO fine-tuning has many advantages over RLHF, such as requiring less
computation, being easier to implement, and is generally more stable with less
hyperparameters~\cite{rafailov2023direct}. However, some works still find that
RL fine-tuning can outperform DPO for certain tasks and
datasets~\cite{wang2023helpsteer}. Thus, we adapt the DPO approach to compare
it with RLPF. We propose Direct Performance Alignment (DPA), an adaptation of
the training procedure and loss function from~\cite{rafailov2023direct} that
takes into account performance, to fine-tune an LLM to generate faster code.
The proposed loss function in DPA is shown in~\Cref{eq:dpa-loss}.

\vspace{1.25em}
\begin{multline}\label{eq:dpa-loss}
    \mathcal{L}_d = 
    - \log
    \sigma
    \bigg(
        \beta
        \log
        \eqnmarkbox[PineGreen]{rdf}{\frac{
            \pi^P(d_f \mid p)
        }{
            \pi^S(d_f \mid p)
        }}
        -
        \beta
        \log
        \eqnmarkbox[PineGreen]{rds}{\frac{
            \pi^P(d_s \mid p)
        }{
            \pi^S(d_s \mid p)
        }} \\
        -
        \mu(p, d_f, d_s)
    \bigg)
\end{multline}
\annotatetwo[yshift=0.75em,xshift=-0.25em]{above}{rdf}{rds}{predicted probabilities from fine-tuned model $\pi^P$ and \\ supervised model $\pi^S$ on fast ($d_f$) and slow ($d_s$) code}

\begin{figure}[ht]
    \vspace{-1em}
    \centering
    \includegraphics[width=\columnwidth]{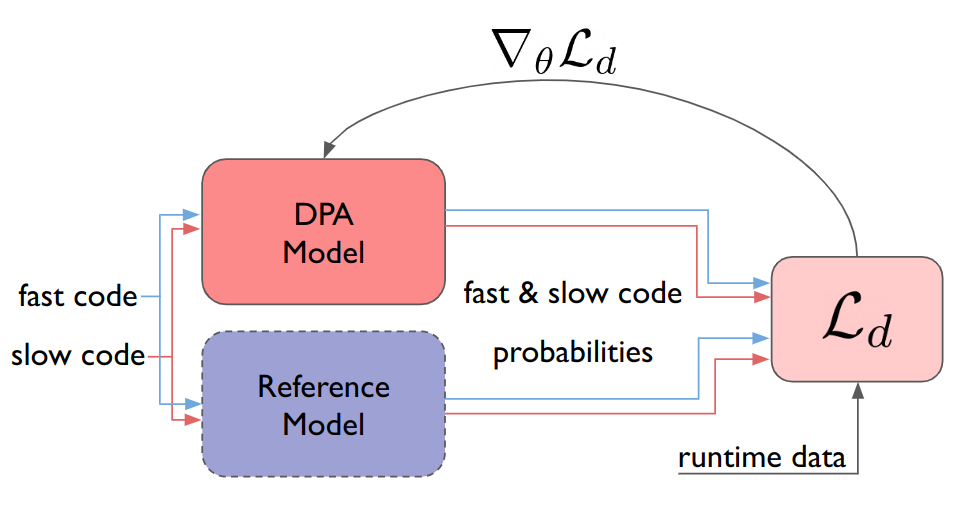}
    \caption{\label{fig:dpa-fine-tuning}
    The DPA fine-tuning process. The model being fine-tuned and a reference
    model are used to generate probabilities for a fast and slow code sample.
    These probabilities, combined with runtime data, are used to compute a loss
    and update the model's parameters.}
\end{figure}

Like with the reward loss in~\Cref{eq:reward-loss}, we utilize the adaptive margin $\mu$
from~\Cref{eq:reward-loss-margin} to scale the loss based on the runtime of the
fast and slow code samples. This loss function can be used to fine-tune a base
LLM to generate faster code without using reinforcement learning. To compute the
loss, we need to get model predictions for a fast and slow code pair for both the
model being fine-tuned and a base reference model (the supervised model). Then
the loss from~\Cref{eq:dpa-loss} is used to update the weights of the model
being fine-tuned. This process is iteratively repeated for a number of
iterations $T$ or until the model converges. This DPA fine-tuning process is
portrayed in~\Cref{fig:dpa-fine-tuning}.

\vspace{0.07in}
\noindent{\bf DPA Fine-Tuning Evaluation Metric}:
The success of DPA fine-tuning can be measured
using a similar accuracy metric to the reward model from RLPF. Since we do not
have a direct reward signal like in~\Cref{eq:reward-loss}, we can instead
measure how often the difference in log probabilities between the fine-tuned
model and the supervised model for the fast code, i.e. $\log\frac{\pi^P(d_f
\mid p)}{\pi^S(d_f \mid p)}$ is greater than the log probability difference for
the slow code, i.e. $\log\frac{\pi^P(d_s \mid p)}{\pi^S(d_s \mid p)}$. This is
shown in~\Cref{eq:dpa-accuracy}.

\begin{equation}\label{eq:dpa-accuracy}
    \text{acc}_{\mathrm{dpa}}(\mathcal{X}) = \frac{1}{|\mathcal{X}|} \sum_{(p, d_f, d_s) \in \mathcal{X}} 
    \mathbbm{1}\left[ \frac{\pi^{P}(d_f \mid p)}{\pi^S(d_f \mid p)} > \frac{\pi^{P}(d_s \mid p)}{\pi^S(d_s \mid p)} \right]
\end{equation}

\section{Evaluation Tasks}
\label{sec:eval}
It is important to quantify how well the models do on downstream tasks after
fine-tuning. In this section we present two different tasks, code generation and
optimization, to evaluate how well the training methodologies in~\Cref{sec:rl}
improved the LLMs ability to generate fast code. We further detail an ablation
study to motivate the use of synthetic data.

\subsection{Code Generation}\label{sec:eval-code-gen}

To evaluate the ability of the models to generate fast code, we utilize two
sets of coding problems. The first is a subset of 100 coding contest problems
from the CodeContests dataset~\cite{li2022competition}
(see~\Cref{sec:perf-dataset}) that were removed from the training set. We can
provide the model with the problem statement and use it to write a solution to
the problem. We can then run the code and measure both its {\it correctness}
and {\it performance}. Correctness can easily be tested using the unit tests
provided with the problems.

In addition to the coding contest problems, we also evaluate the models on the
ParEval benchmark~\cite{nichols:arxiv2024}, which is a collection of parallel
code generation problems for evaluating the ability of LLMs to generate correct
and efficient parallel code. We narrow our focus to a subset of 180 problems,
namely the serial, OpenMP~\cite{OpenMP4}, and MPI~\cite{snir1998mpi} problems.
We include OpenMP and MPI problems to evaluate the models' ability to generate
fast parallel code.
The problems in ParEval range a wide variety of domains, such as linear
algebra, graph algorithms, sorting, etc. The problems are designed to be
challenging and require the generation of efficient code. The ParEval benchmark
provides a great way to test the LLMs on problems unlike what is in their
training data (coding contests).

\vspace{0.07in}
\noindent{\bf Code Generation Evaluation Metrics}: 
We evaluate the generated code on two metrics: {\it correctness} and {\it
performance}. To study correctness we adopt the popular pass@$k$ metric from
Chen et al~\cite{codex-copilot-short-author}. This metric measures the
probability that if an LLM is given $k$ attempts to write a correct solution, it
will succeed. Equation~\ref{eq:pass-at-k} shows how this value can be estimated
using $N$ generated samples from an LLM. Typically the average pass@$k$ over a
set of prompts is reported and, as LLMs have progressed, only the
pass@1 value is reported. We refer the reader
to~\cite{codex-copilot-short-author} for further discussion of pass@k.

\vspace{1.0em}
\begin{equation}\label{eq:pass-at-k}
    \text{pass@}k =
    \frac{1}{\lvert \eqnmarkbox[MidnightBlue]{P1}{P}\rvert}
    \sum_{p\in \eqnmarkbox[MidnightBlue]{P2}{P}}
    \left[
        1 -
        \binom{
            \eqnmarkbox[WildStrawberry]{N1}{N} -
            \eqnmarkbox[OliveGreen]{cp}{c_p}
        }{
            k
        }
        /
        \binom{
            \eqnmarkbox[WildStrawberry]{N2}{N}
        }{
            k
        }
    \right]
\end{equation}
\annotatetwo[yshift=1em]{above}{N1}{N2}{Number of samples generated per prompt}
\annotatetwo[yshift=-1em]{below}{P1}{P2}{Set of prompts}
\annotate[yshift=-2.3em]{below,right}{cp}{Number of correct\\samples for prompt $p$}
\vspace{1.5em}

To evaluate the performance of the generated code, we use the speedup$_n@k$
metric introduced by Nichols et al~\cite{nichols:arxiv2024}. This metric
measures the expected max speedup over a baseline implementation if the LLM is
given $k$ attempts to write a solution. The speedup$_n@k$ metric is defined in
Equation~\ref{eq:speedup-at-k}. We refer the reader to~\cite{nichols:arxiv2024}
for a complete derivation of this metric. For the coding contest problems, we
use the median submission runtime as the baseline. For the ParEval problems, we
use the baselines provided by the benchmark.

\vspace{1.0em}
\begin{equation}\label{eq:speedup-at-k}
    \textrm{speedup}_n@k =
    \frac{1}{\lvert P\rvert}
    \sum_{p\in P}
    \sum_{j=1}^N
    \frac{
        \binom{
            j-1
        }{
            k-1
        }
    }{
        \binom{
            N
        }{
            k
        }
    }
    \frac{
        \eqnmarkbox[SeaGreen]{Tsp}{T^*_p}
    }{
        \eqnmarkbox[Melon]{Tpj}{T_{p,j,n}}
    }
\end{equation}
\annotate[yshift=-0.5em]{below,left}{Tpj}{runtime of sample $j$ of prompt $p$
    on $n$ processors}
\annotate[yshift=1em]{above,left}{Tsp}{runtime of baseline for prompt $p$}
\vspace{1.0em}

\subsection{Code Optimization}\label{sec:eval-code-opt}

In addition to generating code, we also evaluate the ability of the models to
optimize existing code. This is accomplished by providing a code snippet and
instructing the model to generate an optimized version of it. To evaluate this
task we use the functions in the PolyBench benchmark suite~\cite{polybench}.
This is comprised of 30 unique kernels that are typically used to test compiler
optimizations and auto-tuning tools. We utilize the kernels by providing the
existing kernel implementation to the LLM and instructing it to generate an
optimized implementation. We can then evaluate the correctness and performance
of the generated code.

\vspace{0.07in}
\noindent{\bf Code Optimization Evaluation Metrics}: 
We evaluate the generated code on the same metrics as the code generation task:
{\it correctness} and {\it performance}. We use the same pass@$k$ metric
(\Cref{eq:pass-at-k}) to evaluate correctness. To evaluate performance, we use
speedup$_n@k$ (\Cref{eq:speedup-at-k}), except with the baseline being the
runtime of the original kernel.

\subsection{Synthetic Data Ablation Study}\label{sec:eval-synthetic-ablation}

Finally, we test our hypothesis that training on synthetic data helps the
models' ability to generalize and prevents it from over-fitting to code contest
data. To accomplish this we train the models exclusively on the code contests
dataset $\mathcal{D}_c$ without any of the synthetic dataset $\mathcal{D}_s$.
We then evaluate the models on the code generation (\Cref{sec:eval-code-gen})
and code optimization (\Cref{sec:eval-code-opt}) tasks. We compute the same
pass@$k$ and speedup$_n@k$ metrics and compare the impact of the synthetic data
on the models' performance. Of most interest is the performance on the ParEval
and PolyBench benchmarks, as these are the most different from the training
data.

\section{Experimental Setup}
\label{sec:setup}
Using the large performance dataset $\mathcal{D}$ from~\Cref{sec:data} and the
training methodology introduced in~\Cref{sec:rl}, we can now fine-tune LLMs to
generate faster code. Once fine-tuned, these models can then be evaluated on
the benchmarks detailed in~\Cref{sec:eval}. This section details the base
models for fine-tuning, the data subsets for each fine-tuning task, how we
implement the fine-tuning process, and the experimental setup used to evaluate
the fine-tuned models.

\subsection{Base Model for Fine-Tuning}\label{sec:rl-model}

Each of the training methodologies introduced in~\Cref{sec:rl} begins with a
base LLM that has already been trained and fine-tunes it further. We select the
Deepseek-Coder 6.7B model~\cite{guo2024deepseekcoder} as the base for the
supervised fine-tuning (\Cref{sec:supervised}). This model is a 6.7B parameter
code LLM released by Deepseek-AI that is trained on 2T tokens comprised of
mostly code with a context length of 16k tokens. We select this model due to its
good performance on code generation tasks~\cite{bigcode_leaderboard} and due to
other works finding it a better base model for fine-tuning than the popular
CodeLlama models~\cite{wei2023magicoder}. Furthermore, its 6.7B parameter size
makes it tractable for end-users to use it to generate code themselves on
consumer hardware. While Deepseek-Coder is a strong base model for our studies,
the proposed fine-tuning methodologies can be applied to any existing code LLM.

For the remaining two fine-tuning methods, RLPF and DPA, we use the supervised
fine-tuned deepseek model as the base. This is in line with the methodologies
in~\cite{ouyang2022training, rafailov2023direct} and ensures that the model
being aligned is within the distribution of the text data it is trying to model
(i.e. instruction prompts as shown in~\Cref{lst:sft-prompt}). Additionally, we
use Deepseek-Coder 6.7B as the base for the reward model. The final set of
models used for comparison is shown in~\Cref{tab:models-setup}.

\begin{table}[h]
    \centering
    \caption{Models used for comparison in this paper. Deepseek-Coder-6.7B~\cite{guo2024deepseekcoder} is the base model we use in our fine-tuning methodologies.}
    \label{tab:models-setup}
    \begin{tabular}{@{}llc@{}}
    \toprule
    \textbf{Model Name} & \textbf{Description}            & \textbf{\begin{tabular}[c]{@{}l@{}}Fine-Tuning \\ Methodology\end{tabular}} \\ \midrule
    DS                  & Deepseek-Coder 6.7B base model  & ---                                                                         \\
    DS+SFT              & DS after supervised fine-tuning & \Cref{sec:supervised}                                                       \\
    DS+RLPF             & DS+SFT after RLPF fine-tuning   & \Cref{sec:rlpf}                                                             \\
    DS+DPA              & DS+SFT after DPA fine-tuning    & \Cref{sec:dpa}                                                              \\ \bottomrule
    \end{tabular}
    \vspace{-1.5em}
\end{table}

\subsection{Data Setup}\label{sec:data-setup}

We fine-tune the LLMs using the dataset $\mathcal{D}$ from~\Cref{sec:data}. We
set aside 100 contests from the CodeContests dataset for the code generation
evaluation task. The dataset is further split into smaller datasets for each
fine-tuning task. The supervised fine-tuning dataset,
$\mathcal{D}_{\text{SFT}}$, is comprised of 40\% of the full dataset,
$\mathcal{D}$, and the remaining 60\% is used for the reinforcement fine-tuning
dataset, $\mathcal{D}_{\text{RLPF}}$, and the direct performance alignment
dataset, $\mathcal{D}_{\text{DPA}}$. These two datasets can be the same since
the alignment fine-tuning tasks are disjoint. The $\mathcal{D}_{\text{RLPF}}$
dataset is further split into 66\% for the reward model dataset,
$\mathcal{D}_{\text{REWARD}}$, and 33\% for the reinforcement learning dataset,
$\mathcal{D}_{\text{RL}}$. During each fine-tuning stage we set aside 5\% of the
respective dataset for evaluation (i.e. 5\% of $\mathcal{D}_{\text{REWARD}}$ is
set aside to calculate the reward model accuracy after training). All of the
dataset splits are stratified so that the proportion of code contest to
synthetic data is equal to the original dataset.

When creating prompt, fast code, and slow code triplets $(p, d_f, d_s)$ from
$\mathcal{D}_c$ for RLPF and DPA fine-tuning, we select $d_f$ randomly from the
top 5 fastest solutions. We then select $d_s$ from the slowest 50\% of the
solutions. Additionally, a random 5\% subset of slow solutions are replaced
with an incorrect solution. This is to ensure that the model is not just
learning to generate fast code, but also to avoid generating incorrect code. We
directly use the fast and slow code pairs from $\mathcal{D}_s$ to directly form
the triplet.

\subsection{Fine-Tuning Setup}\label{sec:rl-setup}

In order to implement the fine-tuning we extend the TRL Python
library~\cite{vonwerra2022trl} which is built on top of the popular transformers
library~\cite{huggingface}. TRL provides existing implementations of RLHF and
DPO, which we modify to use our custom rewards, loss function, and datasets. We
fine-tune the models on a single node with four 80GB A100 GPUs and two AMD EPYC
7763 CPUs.

\subsubsection{Supervised Fine-Tuning Hyperparameters}\label{sec:supervised-setup}

We fine-tune the supervised model for three epochs over the
$\mathcal{D}_{\text{SFT}}$ dataset. We use bfloat16 precision and a global
batch size of 64 (1 sample per GPU and 16 gradient accumulation steps). To
fine-tune in parallel we make use of the PyTorch fully sharded data parallelism
(FSDP) implementation~\cite{fsdp}, which shards model parameters across ranks
to save memory. Furthermore, we fine-tune with the Adam
optimizer~\cite{KingmaAdam2014} and an initial learning rate of
$1.41\times10^{-5}$.

\subsubsection{Reward Model Fine-Tuning Hyperparameters}\label{sec:reward-setup}

The reward model is fine-tuned with the same hyperparameters as the supervised
model (\Cref{sec:supervised-setup}), except it is fine-tuned for only one epoch
over the $\mathcal{D}_{\text{REWARD}}$ dataset.  We use a max margin of
$\lambda=3$ for the margin function $\mu(p,d_f,d_s)$. 

\subsubsection{RLPF Fine-Tuning Hyperparameters}\label{sec:rlpf-setup}

We fine-tune the RLPF model for four PPO epochs over the
$\mathcal{D}_{\text{RL}}$ dataset. We use a global batch size of four and a
learning rate of $1.41\times10^{-5}$. The KL regularization coefficient is
initialized to $\gamma = 0.1$. When sampling outputs from the fine-tuned and
reference model we follow best conventions~\cite{vonwerra2022trl} and use
sampling with a top-$k$ of 0 and a top-$p$ of 1.0.

\subsubsection{DPA Fine-Tuning Hyperparameters}\label{sec:dpa-setup}

The DPA model is fine-tuned for 1 epoch over the $\mathcal{D}_{\text{DPA}}$
dataset with a global batch size of four. We employ a learning rate of
$1\times10^{-7}$ in the AdamW optimizer~\cite{adamw}. Additionally, we found a
value of $\beta = 0.6$ to be most stable for training.

\subsection{Evaluation Setup}\label{sec:eval-setup}

For the code generation tasks we use each of the LLMs to generate code for the
prompts in the evaluation subset of $\mathcal{D}_c$ and ParEval. We generate 20
samples per prompt with a temperature of 0.2 and a top-$p$ of 0.95 following
standard practices LLM code benchmarks~\cite{mbpp,nichols:arxiv2024}. For the
optimization task we similarly generate 20 optimized versions of each kernel in
the PolyBench benchmark suite~\cite{polybench} using each of the fine-tuned
LLMs.

The generated code is run on a single AMD EPYC 7763 CPU. For the ParEval OpenMP
tests we report results on 8 cores and we use 512 ranks for the MPI tests. We
make use of the existing tests in the CodeContests dataset and ParEval to record
the correctness and runtime of the generated code. For the optimized PolyBench
kernels we test correctness and runtime against the original kernel
implementations. All runtimes are averaged over 5 runs.

\section{Results}
\label{sec:results}
With the fine-tuned models from~\Cref{sec:rl} we can now evaluate their code
generation capabilities on the tasks described in~\Cref{sec:eval}. In this
section we present the results from the fine-tuning process and the evaluation
tasks.

\subsection{Fine-Tuning Results}\label{sec:results-fine-tuning}

We record the fine-tuning metrics on the 5\% evaluation datasets at the end of
each fine-tuning step. The DS+SFT model yields an evaluation perplexity of 1.62.
It is generally difficult to reason about specific perplexity values, but values
near 1 show a strong ability to model the underlying text distribution. Since
perplexity is the exponential of cross-entropy (see~\Cref{eq:perplexity}) a
perplexity value of 1.62 means that the cross-entropy between predicted
probabilities is $\approx 0.48$.

The RLPF reward model achieves a final evaluation accuracy of 93\% after one
epoch of training calculated using~\Cref{eq:reward-accuracy}. This means that
in 93\% of samples the model assigns a higher reward signal to faster code than
slower code. This is a strong result as the success of RL-based LLM fine-tuning
is highly dependent on the quality of the reward model~\cite{wang2024secrets}.
Using this reward model the DS+RLPF model is then able to achieve a mean reward
of 1.8 and a KL divergence of 0.29. This means that DP-RLPF is getting a
positive mean reward, while maintaining a similar distribution to the original
model.

Finally, we see that the DS+DPA model achieves an evaluation accuracy of 87\%
calculated as show in~\Cref{eq:dpa-accuracy}. This is not quite as high as the
RLPF reward model, but is still a strong result. The log-probability difference
between DS+DPA and the reference model for fast code samples is greater than
the log-probability difference for slow code samples in 87\% of the evaluation
dataset.

\subsection{Code Generation Results}\label{sec:results-code-generation}

\begin{figure}[h]
    \vspace{-1em}
    \centering
    \includegraphics[width=\linewidth]{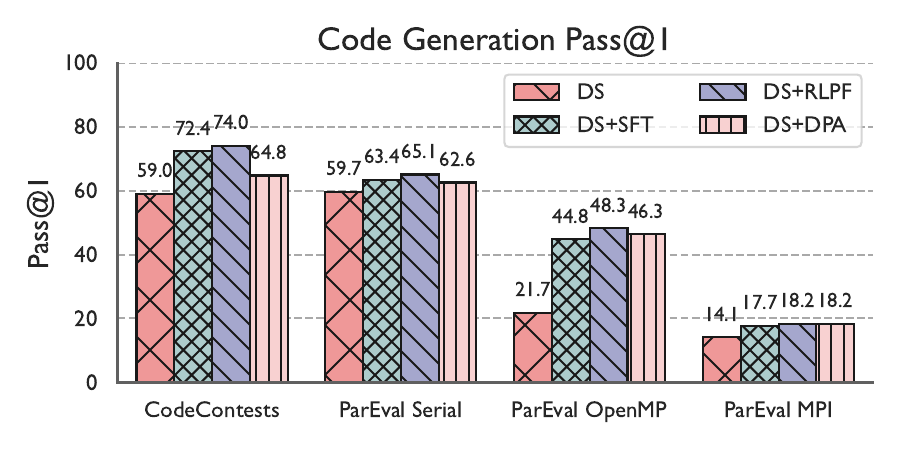}
    \caption{\label{fig:code-generation-pass-1}
        Correctness results for each model on the code generation
        tasks. Each of the fine-tuned models shows an improvement in correctness
        over the baseline model with the DS+RLPF model showing the most
        improvement. }
\end{figure}

\begin{figure}[h]
    \centering
    \includegraphics[width=\linewidth]{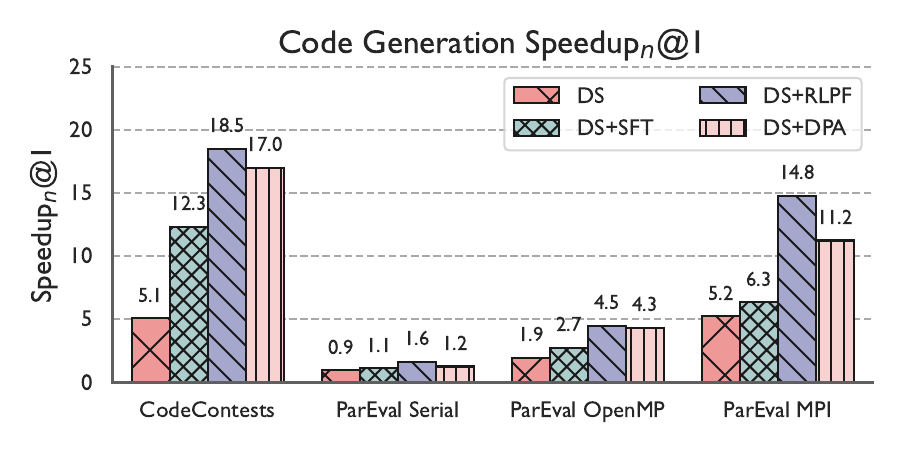}
    \caption{\label{fig:code-generation-speedup-1}
        Speedup results for each fine-tuned model on the code generation tasks.
        OpenMP runtimes are on 8 cores and MPI runtimes are on 512 ranks. The
        DS+RLPF model is the best performing model across all benchmarks. }
\end{figure}

\Cref{fig:code-generation-pass-1,fig:code-generation-speedup-1} show the
correctness and performance results of each fine-tuned model on the code
generation tasks. We see a promising trend in pass@1 scores
in~\Cref{fig:code-generation-pass-1} where the fine-tuned models improve in
correctness over the baseline model. The DS+RLPF model shows the most
improvement across all tasks. These improvements can be attributed to training
over more data and, in the case of the RLPF and DPA models, using incorrect
samples as negative rewards. Improving the correctness of the models is a strong
results considering that the primary goal of this work is to improve the
performance while keeping the correctness levels the same.

\Cref{fig:code-generation-speedup-1} further details the speedup results for
each fine-tuned model. We present the speedup results for OpenMP on 8 cores and
MPI on 512 ranks with a sequential implementation as the baseline. Across all
four benchmarks DS+RLPF produces faster code than the other three models. In the
case of the code contests and ParEval serial problems, the speedup$_1$@1 value
is easy to interpret. For instance, in the case of the serial ParEval problems,
DS+RLPF generates code with an expected max speedup of 1.6x over the sequential
baseline. We see the same order of model performance across all the benchmarks
with DS+RLPF performing the best, followed by DS+DPA, DS+SFT, and DS.

\subsection{Code Optimization Results}\label{sec:results-code-optimization}

\begin{figure}[h]
    \includegraphics[width=0.48\linewidth]{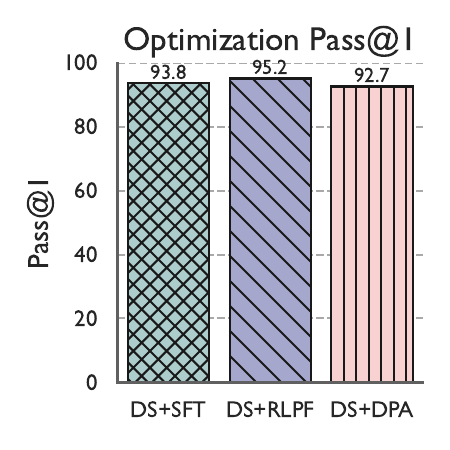}
    \includegraphics[width=0.48\linewidth]{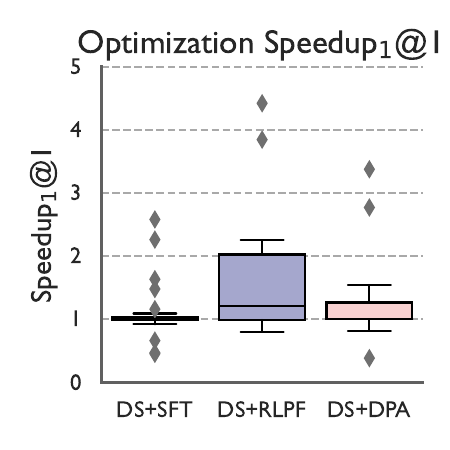}
    \caption{\label{fig:optimization}
    pass@1 (left) and speedup$_1$@1 (right) results for optimizing the PolyBench
    kernels. The distribution of speedup$_1$@1 values over the 30 benchmarks is
    shown on the right. The DS+RLPF model has further outliers at 11.6 and 22.4.
    }
\end{figure}

\Cref{fig:optimization} shows the correctness and performance results when using
the fine-tuned models to optimize PolyBench kernels. DS is omitted because it is
only a code completion model and was not trained to optimize code inputs. We
first see that all three fine-tuned models transform the input code to a correct
output code with relatively high accuracy. While provably correct compiler
optimizations may seem more desirable, LLM optimizations can be applied at a
higher level of abstraction and include natural language comments to explain the
transformation to a developer.

We show the distribution of speedup$_1$@1 per PolyBench benchmark
in~\Cref{fig:optimization} rather than an average to highlight the spread of
results. The speedup results show that DS+RLPF is the best performing model. It
is able to produce an expected max speedup greater than 1 in 26 out of the 30
benchmarks. In the case of the 3mm kernel (three matrix multiplies) it is able
to get up to 22.4x expected speedup. Many of the optimizations come from loop
unrolling and/or cache friendly data access patterns. The DS+DPA model is able
to produce faster optimizations than DS+SFT, but is not as strong as DS+RLPF.

\subsection{Synthetic Data Ablation Study Results}\label{sec:results-synthetic-data-ablation-study}

\begin{figure}[h]
    \centering
    \includegraphics[width=\linewidth]{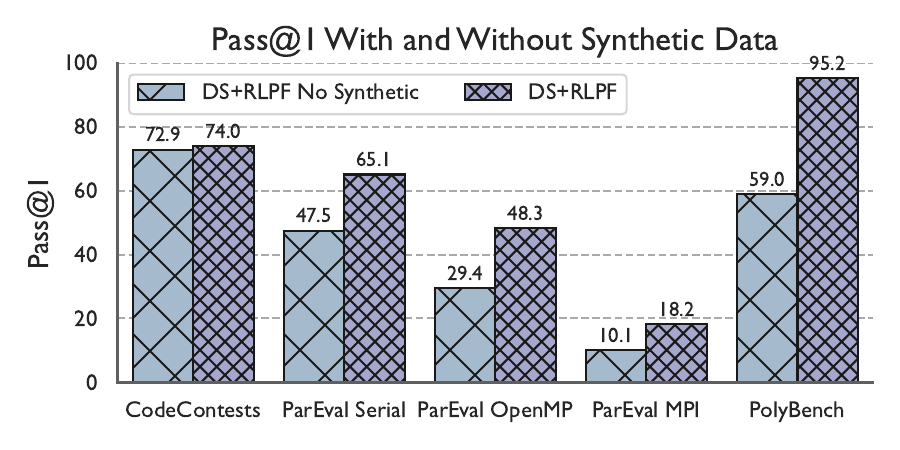}
    \caption{\label{fig:ablation-pass-1}
        pass@1 results for DS+RLPF on each task with and without synthetic data
        in the fine-tuning dataset. For all tasks, the model fine-tuned on
        synthetic data produces correct code at a higher rate. }
\end{figure}

\begin{figure}[h]
    \centering
    \includegraphics[width=\linewidth]{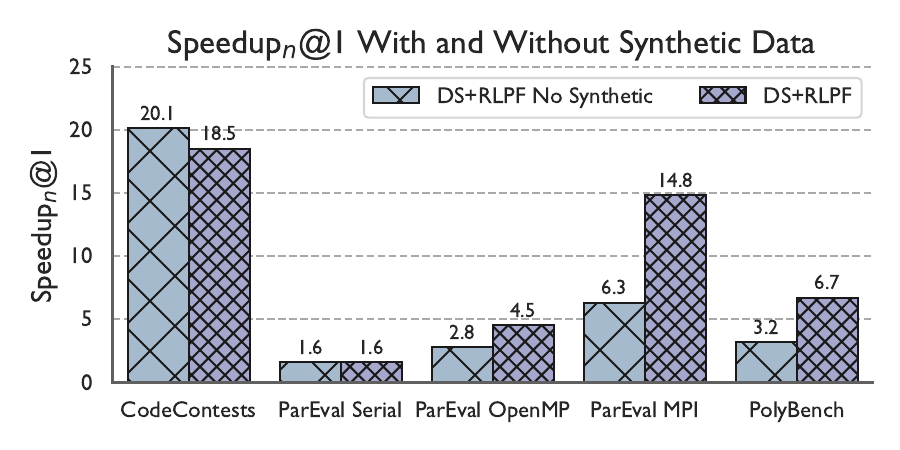}
    \caption{\label{fig:ablation-speedup-1}
        speedup$_n$@1 results for DS+RLPF on each task with and without
        synthetic data in the fine-tuning dataset. For OpenMP, MPI, and
        PolyBench tasks, the model fine-tuned on synthetic data produces faster
        code, while the coding contest and ParEval serial problems show a slight
        decrease or no change in speedup. }
\end{figure}

We further highlight the use of synthetic data in the fine-tuning process
in~\Cref{fig:ablation-pass-1,fig:ablation-speedup-1}. Results for DS+RLPF are
shown since it is the best performing model. We see a general improvement in
both correctness and performance of generated code when incorporating synthetic
data versus fine-tuning on just coding contest data.
The correctness improves for all the benchmarks (\Cref{fig:ablation-pass-1})
and, notably, even improves on the coding contest benchmarks. The broader
synthetic data is able to help the model generalize better even within the 
coding contest domain.

The speedup results in~\Cref{fig:ablation-speedup-1} show that fine-tuning with
synthetic data also helps the models produce faster code. Only in the case of
the coding contests and ParEval serial problems do we see a decrease or no
change in speedup$_n$@1. However, these differences are small. The performance
increases for OpenMP, MPI, and PolyBench are much more significant.
incorporating synthetic performance data into the fine-tuning process has
prevented the models from overfitting code contest data and enabled them to
generalize better to new tasks.

\section{Related Work}
\label{sec:related}
Large Language Models (LLMs), like OpenAI Codex~\cite{codex-copilot-all-author},
CodeLlama~\cite{touvron2023llama}, StarCoder~\cite{li2023starcoder},
WizardCoder~\cite{luo2023wizardcoder},
Phind-CodeLlama~\cite{phind-codellama-34b-v2}, and
DeepSeek~\cite{guo2024deepseekcoder} are revolutionizing how developers approach
their coding tasks. These models are trained on vast datasets that include code
repositories, documentation, high quality programming problems and solutions.
They have shown incredible potential in a variety of software-related tasks
ranging from code completion~\cite{openai2023gpt4, Barke2022GroundedCH,
Dderlein2022PilotingCA, Sarkar2022WhatII, li2022competition, guo2023longcoder},
code refactoring~\cite{white2023chatgpt}, bug detection~\cite{chen2023teaching,
yasunaga2021break, zhang2022repairing,sobania2023analysis},
documentation~\cite{khan2022automatic}, and
testing~\cite{schafer2023empirical,chen2022codet}, among others. 

In HPC, researchers are particularly interested in using LLMs for generating
parallel code~\cite{valerolara2023comparing, chen2023lm4hpc,
ding2023hpc,nichols:arxiv2024,nichols:arxiv2023}. Nichols et
al.~\cite{nichols:arxiv2023} proposed HPCCoder, a model fine-tuned on HPC data,
to generate parallel code, label OpenMP pragmas, and predict performance. Chen
et al. designed LM4HPC~\cite{chen2023lm4hpc} framework to facilitate the
research and development of HPC software and proposed
OMPGPT~\cite{chen2024ompgpt} for generating OpenMP pragmas and data race
detection~\cite{chen2023data}. Despite their popularity, LLMs still struggle at
generating efficient code~\cite{valerolara2023comparing, nichols:arxiv2024}. Our
work addresses this concern by incorporating performance aspects of code in
order to generate efficient code while maintaining its correctness.

While Reinforcement Learning with Human Feedback (RLHF)~\cite{wang2024secrets}
has been shown to be critical for boosting the performance of LLMs by
incorporating human feedback into the reward model~\cite{rafailov2023direct}, it
is not specialized to code and furthermore does not consider performance.
Another work by Mankowitz et al.~\cite{mankowitz2023faster} looked at training a
deep reinforcement learning agent, AlphaDev, to discover sorting algorithms from
scratch that outperformed previously known human benchmarks. However, training
process is limited to a single algorithm at a time and does not fine-tune a
general model that can be used to generate fast code for a variety of problems.
To address this gap and enable LLMs to generate faster versions of code, we
introduced RLPF and DPA to tune LLMs on performance data.

\section{Conclusion}
\label{sec:conclusion}
In this paper, we have explored the idea of fine-tuning large language models
to help them learn code structures and patterns that generally lead to better
performance.  To accomplish this, we first collected a large performance
dataset from coding contests and extended it with synthetically generated
samples to cover a wider distribution of code. We then introduced two novel
fine-tuning methodologies, Reinforcement Learning with Performance Feedback
(RLPF) and Direct Performance Alignment (DPA), that align LLMs with faster code
outputs.  We have demonstrated that using such techniques we can incorporate
performance feedback into the fine-tuning of code LLMs. The fine-tuned models
were evaluated on code generation and optimization tasks and shown to increase
the expected performance of generated code over baseline LLMs while maintaining
correctness for both serial and parallel codes.

\bibliographystyle{IEEEtran}
\bibliography{bib/pssg,bib/cite}

\begin{thebibliography}{10}
\providecommand{\url}[1]{#1}
\csname url@samestyle\endcsname
\providecommand{\newblock}{\relax}
\providecommand{\bibinfo}[2]{#2}
\providecommand{\BIBentrySTDinterwordspacing}{\spaceskip=0pt\relax}
\providecommand{\BIBentryALTinterwordstretchfactor}{4}
\providecommand{\BIBentryALTinterwordspacing}{\spaceskip=\fontdimen2\font plus
\BIBentryALTinterwordstretchfactor\fontdimen3\font minus
  \fontdimen4\font\relax}
\providecommand{\BIBforeignlanguage}[2]{{%
\expandafter\ifx\csname l@#1\endcsname\relax
\typeout{** WARNING: IEEEtran.bst: No hyphenation pattern has been}%
\typeout{** loaded for the language `#1'. Using the pattern for}%
\typeout{** the default language instead.}%
\else
\language=\csname l@#1\endcsname
\fi
#2}}
\providecommand{\BIBdecl}{\relax}
\BIBdecl

\bibitem{codex-copilot-short-author}
M.~Chen and et~al, ``Evaluating large language models trained on code,'' 2021.

\bibitem{Richter2022CanWL}
C.~Richter and H.~Wehrheim, ``Can we learn from developer mistakes? learning to
  localize and repair real bugs from real bug fixes,'' \emph{ArXiv}, vol.
  abs/2207.00301, 2022.

\bibitem{Kharkar2022LearningTR}
A.~Kharkar, R.~Z. Moghaddam, M.~Jin, X.~Liu, X.~Shi, C.~B. Clement, and
  N.~Sundaresan, ``Learning to reduce false positives in analytic bug
  detectors,'' \emph{2022 IEEE/ACM 44th International Conference on Software
  Engineering (ICSE)}, pp. 1307--1316, 2022.

\bibitem{Ahmad2020ATA}
W.~U. Ahmad, S.~Chakraborty, B.~Ray, and K.-W. Chang, ``A transformer-based
  approach for source code summarization,'' \emph{ArXiv}, vol. abs/2005.00653,
  2020.

\bibitem{Haque2022SemanticSM}
S.~Haque, Z.~Eberhart, A.~Bansal, and C.~McMillan, ``Semantic similarity
  metrics for evaluating source code summarization,'' \emph{2022 IEEE/ACM 30th
  International Conference on Program Comprehension (ICPC)}, pp. 36--47, 2022.

\bibitem{Gu2022AssembleFM}
J.~Gu, P.~Salza, and H.~C. Gall, ``Assemble foundation models for automatic
  code summarization,'' \emph{2022 IEEE International Conference on Software
  Analysis, Evolution and Reengineering (SANER)}, pp. 935--946, 2022.

\bibitem{Ahmed2022LearningCS}
T.~Ahmed and P.~Devanbu, ``Learning code summarization from a small and local
  dataset,'' \emph{ArXiv}, vol. abs/2206.00804, 2022.

\bibitem{nichols:arxiv2023}
D.~Nichols, A.~Marathe, H.~Menon, T.~Gamblin, and A.~Bhatele, ``Modeling
  parallel programs using large language models,'' 2023.

\bibitem{nichols:arxiv2024}
D.~Nichols, J.~H. Davis, Z.~Xie, A.~Rajaram, and A.~Bhatele, ``Can large
  language models write parallel code?'' 2024.

\bibitem{valerolara2023comparing}
P.~Valero-Lara, A.~Huante, M.~A. Lail, W.~F. Godoy, K.~Teranishi,
  P.~Balaprakash, and J.~S. Vetter, ``Comparing llama-2 and gpt-3 llms for hpc
  kernels generation,'' 2023.

\bibitem{ouyang2022training}
L.~Ouyang, J.~Wu, X.~Jiang, D.~Almeida, C.~L. Wainwright, P.~Mishkin, C.~Zhang,
  S.~Agarwal, K.~Slama, A.~Ray, J.~Schulman, J.~Hilton, F.~Kelton, L.~Miller,
  M.~Simens, A.~Askell, P.~Welinder, P.~Christiano, J.~Leike, and R.~Lowe,
  ``Training language models to follow instructions with human feedback,''
  2022.

\bibitem{rafailov2023direct}
R.~Rafailov, A.~Sharma, E.~Mitchell, S.~Ermon, C.~D. Manning, and C.~Finn,
  ``Direct preference optimization: Your language model is secretly a reward
  model,'' 2023.

\bibitem{li2023starcoder}
R.~Li, L.~B. Allal, Y.~Zi, N.~Muennighoff, D.~Kocetkov, C.~Mou, M.~Marone,
  C.~Akiki, J.~Li, J.~Chim, Q.~Liu, E.~Zheltonozhskii, T.~Y. Zhuo, T.~Wang,
  O.~Dehaene, M.~Davaadorj, J.~Lamy-Poirier, J.~Monteiro, O.~Shliazhko,
  N.~Gontier, N.~Meade, A.~Zebaze, M.-H. Yee, L.~K. Umapathi, J.~Zhu,
  B.~Lipkin, M.~Oblokulov, Z.~Wang, R.~Murthy, J.~Stillerman, S.~S. Patel,
  D.~Abulkhanov, M.~Zocca, M.~Dey, Z.~Zhang, N.~Fahmy, U.~Bhattacharyya, W.~Yu,
  S.~Singh, S.~Luccioni, P.~Villegas, M.~Kunakov, F.~Zhdanov, M.~Romero,
  T.~Lee, N.~Timor, J.~Ding, C.~Schlesinger, H.~Schoelkopf, J.~Ebert, T.~Dao,
  M.~Mishra, A.~Gu, J.~Robinson, C.~J. Anderson, B.~Dolan-Gavitt,
  D.~Contractor, S.~Reddy, D.~Fried, D.~Bahdanau, Y.~Jernite, C.~M. Ferrandis,
  S.~Hughes, T.~Wolf, A.~Guha, L.~von Werra, and H.~de~Vries, ``Starcoder: may
  the source be with you!'' 2023.

\bibitem{li2022competition}
Y.~Li, D.~Choi, J.~Chung, N.~Kushman, J.~Schrittwieser, R.~Leblond, T.~Eccles,
  J.~Keeling, F.~Gimeno, A.~Dal~Lago, T.~Hubert, P.~Choy,
  C.~de~Masson~d'Autume, I.~Babuschkin, X.~Chen, P.-S. Huang, J.~Welbl,
  S.~Gowal, A.~Cherepanov, J.~Molloy, D.~Mankowitz, E.~Sutherland~Robson,
  P.~Kohli, N.~de~Freitas, K.~Kavukcuoglu, and O.~Vinyals, ``Competition-level
  code generation with alphacode,'' \emph{arXiv preprint arXiv:2203.07814},
  2022.

\bibitem{transformer}
\BIBentryALTinterwordspacing
A.~Vaswani, N.~Shazeer, N.~Parmar, J.~Uszkoreit, L.~Jones, A.~N. Gomez,
  L.~Kaiser, and I.~Polosukhin, ``Attention is all you need,'' \emph{CoRR},
  vol. abs/1706.03762, 2017. [Online]. Available:
  \url{http://arxiv.org/abs/1706.03762}
\BIBentrySTDinterwordspacing

\bibitem{guo2024deepseekcoder}
D.~Guo, Q.~Zhu, D.~Yang, Z.~Xie, K.~Dong, W.~Zhang, G.~Chen, X.~Bi, Y.~Wu,
  Y.~K. Li, F.~Luo, Y.~Xiong, and W.~Liang, ``Deepseek-coder: When the large
  language model meets programming -- the rise of code intelligence,'' 2024.

\bibitem{wei2023magicoder}
Y.~Wei, Z.~Wang, J.~Liu, Y.~Ding, and L.~Zhang, ``Magicoder: Source code is all
  you need,'' \emph{arXiv preprint arXiv:2312.02120}, 2023.

\bibitem{holtzman:iclr2020}
\BIBentryALTinterwordspacing
A.~Holtzman, J.~Buys, L.~Du, M.~Forbes, and Y.~Choi, ``The curious case of
  neural text degeneration,'' in \emph{International Conference on Learning
  Representations}, 2020. [Online]. Available:
  \url{https://openreview.net/forum?id=rygGQyrFvH}
\BIBentrySTDinterwordspacing

\bibitem{schulman2017proximal}
J.~Schulman, F.~Wolski, P.~Dhariwal, A.~Radford, and O.~Klimov, ``Proximal
  policy optimization algorithms,'' 2017.

\bibitem{aizu}
``Aizu,'' \url{https://judge.u-aizu.ac.jp/onlinejudge/}.

\bibitem{atcoder}
``Atcoder,'' \url{https://atcoder.jp/}.

\bibitem{codechef}
``Codechef,'' \url{https://www.codechef.com/}.

\bibitem{codeforces}
``Codeforces,'' \url{https://codeforces.com/}.

\bibitem{hackerearth}
``Hackerearth,'' \url{https://www.hackerearth.com/}.

\bibitem{zheng2023judging}
L.~Zheng, W.-L. Chiang, Y.~Sheng, S.~Zhuang, Z.~Wu, Y.~Zhuang, Z.~Lin, Z.~Li,
  D.~Li, E.~P. Xing, H.~Zhang, J.~E. Gonzalez, and I.~Stoica, ``Judging
  llm-as-a-judge with mt-bench and chatbot arena,'' 2023.

\bibitem{Gilardi_2023}
\BIBentryALTinterwordspacing
F.~Gilardi, M.~Alizadeh, and M.~Kubli, ``Chatgpt outperforms crowd workers for
  text-annotation tasks,'' \emph{Proceedings of the National Academy of
  Sciences}, vol. 120, no.~30, Jul. 2023. [Online]. Available:
  \url{http://dx.doi.org/10.1073/pnas.2305016120}
\BIBentrySTDinterwordspacing

\bibitem{he2023annollm}
X.~He, Z.~Lin, Y.~Gong, A.-L. Jin, H.~Zhang, C.~Lin, J.~Jiao, S.~M. Yiu,
  N.~Duan, and W.~Chen, ``Annollm: Making large language models to be better
  crowdsourced annotators,'' 2023.

\bibitem{benallal2024cosmopedia}
\BIBentryALTinterwordspacing
L.~Ben~Allal, A.~Lozhkov, G.~Penedo, T.~Wolf, and L.~von Werra, ``Cosmopedia,''
  2024. [Online]. Available:
  \url{https://huggingface.co/datasets/HuggingFaceTB/cosmopedia}
\BIBentrySTDinterwordspacing

\bibitem{geminishort2023gemini}
G.~Team, ``Gemini: A family of highly capable multimodal models,'' 2023.

\bibitem{Kocetkov2022TheStack}
D.~Kocetkov, R.~Li, L.~Ben~Allal, J.~Li, C.~Mou, C.~Muñoz~Ferrandis,
  Y.~Jernite, M.~Mitchell, S.~Hughes, T.~Wolf, D.~Bahdanau, L.~von Werra, and
  H.~de~Vries, ``The stack: 3 tb of permissively licensed source code,''
  \emph{Preprint}, 2022.

\bibitem{mbpp}
\BIBentryALTinterwordspacing
J.~Austin, A.~Odena, M.~I. Nye, M.~Bosma, H.~Michalewski, D.~Dohan, E.~Jiang,
  C.~J. Cai, M.~Terry, Q.~V. Le, and C.~Sutton, ``Program synthesis with large
  language models,'' \emph{CoRR}, vol. abs/2108.07732, 2021. [Online].
  Available: \url{https://arxiv.org/abs/2108.07732}
\BIBentrySTDinterwordspacing

\bibitem{openai2023gpt4}
OpenAI, ``Gpt-4 technical report,'' 2023.

\bibitem{naveed2024comprehensive}
H.~Naveed, A.~U. Khan, S.~Qiu, M.~Saqib, S.~Anwar, M.~Usman, N.~Akhtar,
  N.~Barnes, and A.~Mian, ``A comprehensive overview of large language
  models,'' 2024.

\bibitem{ziegler2020finetuning}
D.~M. Ziegler, N.~Stiennon, J.~Wu, T.~B. Brown, A.~Radford, D.~Amodei,
  P.~Christiano, and G.~Irving, ``Fine-tuning language models from human
  preferences,'' 2020.

\bibitem{bai2022training}
Y.~Bai, A.~Jones, K.~Ndousse, A.~Askell, A.~Chen, N.~DasSarma, D.~Drain,
  S.~Fort, D.~Ganguli, T.~Henighan, N.~Joseph, S.~Kadavath, J.~Kernion,
  T.~Conerly, S.~El-Showk, N.~Elhage, Z.~Hatfield-Dodds, D.~Hernandez, T.~Hume,
  S.~Johnston, S.~Kravec, L.~Lovitt, N.~Nanda, C.~Olsson, D.~Amodei, T.~Brown,
  J.~Clark, S.~McCandlish, C.~Olah, B.~Mann, and J.~Kaplan, ``Training a
  helpful and harmless assistant with reinforcement learning from human
  feedback,'' 2022.

\bibitem{wang2024secrets}
B.~Wang, R.~Zheng, L.~Chen, Y.~Liu, S.~Dou, C.~Huang, W.~Shen, S.~Jin, E.~Zhou,
  C.~Shi, S.~Gao, N.~Xu, Y.~Zhou, X.~Fan, Z.~Xi, J.~Zhao, X.~Wang, T.~Ji,
  H.~Yan, L.~Shen, Z.~Chen, T.~Gui, Q.~Zhang, X.~Qiu, X.~Huang, Z.~Wu, and
  Y.-G. Jiang, ``Secrets of rlhf in large language models part ii: Reward
  modeling,'' 2024.

\bibitem{jaques2019way}
N.~Jaques, A.~Ghandeharioun, J.~H. Shen, C.~Ferguson, A.~Lapedriza, N.~Jones,
  S.~Gu, and R.~Picard, ``Way off-policy batch deep reinforcement learning of
  implicit human preferences in dialog,'' 2019.

\bibitem{laidlaw2023preventing}
\BIBentryALTinterwordspacing
C.~Laidlaw, S.~Singhal, and A.~Dragan, ``Preventing reward hacking with
  occupancy measure regularization,'' in \emph{ICML Workshop on New Frontiers
  in Learning, Control, and Dynamical Systems}, 2023. [Online]. Available:
  \url{https://openreview.net/forum?id=oiT8js6p3Z}
\BIBentrySTDinterwordspacing

\bibitem{wang2023helpsteer}
Z.~Wang, Y.~Dong, J.~Zeng, V.~Adams, M.~N. Sreedhar, D.~Egert, O.~Delalleau,
  J.~P. Scowcroft, N.~Kant, A.~Swope, and O.~Kuchaiev, ``Helpsteer:
  Multi-attribute helpfulness dataset for steerlm,'' 2023.

\bibitem{OpenMP4}
``{OpenMP Application Program Interface. Version 4.0. July 2013},'' 2013.

\bibitem{snir1998mpi}
\BIBentryALTinterwordspacing
M.~Snir, \emph{MPI--the Complete Reference: The MPI core}, ser. MPI: The
  Complete Reference.\hskip 1em plus 0.5em minus 0.4em\relax Mass, 1998.
  [Online]. Available: \url{https://books.google.com/books?id=x79puJ2YkroC}
\BIBentrySTDinterwordspacing

\bibitem{polybench}
J.~C.~S. Grauer-Gray, ``Polybench,''
  \url{https://web.cs.ucla.edu/~pouchet/software/polybench/}, 2012.

\bibitem{bigcode_leaderboard}
\BIBentryALTinterwordspacing
``Big code models leaderboard - a hugging face space by bigcode,'' 2023.
  [Online]. Available:
  \url{https://huggingface.co/spaces/bigcode/bigcode-models-leaderboard}
\BIBentrySTDinterwordspacing

\bibitem{vonwerra2022trl}
L.~von Werra, Y.~Belkada, L.~Tunstall, E.~Beeching, T.~Thrush, N.~Lambert, and
  S.~Huang, ``Trl: Transformer reinforcement learning,''
  \url{https://github.com/huggingface/trl}, 2020.

\bibitem{huggingface}
\BIBentryALTinterwordspacing
T.~Wolf, L.~Debut, V.~Sanh, J.~Chaumond, C.~Delangue, A.~Moi, P.~Cistac, C.~Ma,
  Y.~Jernite, J.~Plu, C.~Xu, T.~Le~Scao, S.~Gugger, M.~Drame, Q.~Lhoest, and
  A.~M. Rush, ``{Transformers: State-of-the-Art Natural Language
  Processing}.''\hskip 1em plus 0.5em minus 0.4em\relax Association for
  Computational Linguistics, Oct. 2020, pp. 38--45. [Online]. Available:
  \url{https://www.aclweb.org/anthology/2020.emnlp-demos.6}
\BIBentrySTDinterwordspacing

\bibitem{fsdp}
Y.~Zhao, A.~Gu, R.~Varma, L.~Luo, C.-C. Huang, M.~Xu, L.~Wright,
  H.~Shojanazeri, M.~Ott, S.~Shleifer, A.~Desmaison, C.~Balioglu, P.~Damania,
  B.~Nguyen, G.~Chauhan, Y.~Hao, A.~Mathews, and S.~Li, ``Pytorch fsdp:
  Experiences on scaling fully sharded data parallel,'' \emph{Proc. VLDB
  Endow.}, vol.~16, no.~12, p. 3848–3860, aug 2023.

\bibitem{KingmaAdam2014}
\BIBentryALTinterwordspacing
D.~P. Kingma and J.~Ba, ``Adam: {A} method for stochastic optimization,'' in
  \emph{3rd International Conference on Learning Representations, {ICLR} 2015,
  San Diego, CA, USA, May 7-9, 2015, Conference Track Proceedings}, Y.~Bengio
  and Y.~LeCun, Eds., 2015. [Online]. Available:
  \url{http://arxiv.org/abs/1412.6980}
\BIBentrySTDinterwordspacing

\bibitem{adamw}
\BIBentryALTinterwordspacing
I.~Loshchilov and F.~Hutter, ``Fixing weight decay regularization in adam,''
  \emph{CoRR}, vol. abs/1711.05101, 2017. [Online]. Available:
  \url{http://arxiv.org/abs/1711.05101}
\BIBentrySTDinterwordspacing

\bibitem{codex-copilot-all-author}
M.~Chen, J.~Tworek, H.~Jun, Q.~Yuan, H.~P. de~Oliveira~Pinto, J.~Kaplan,
  H.~Edwards, Y.~Burda, N.~Joseph, G.~Brockman, A.~Ray, R.~Puri, G.~Krueger,
  M.~Petrov, H.~Khlaaf, G.~Sastry, P.~Mishkin, B.~Chan, S.~Gray, N.~Ryder,
  M.~Pavlov, A.~Power, L.~Kaiser, M.~Bavarian, C.~Winter, P.~Tillet, F.~P.
  Such, D.~Cummings, M.~Plappert, F.~Chantzis, E.~Barnes, A.~Herbert-Voss,
  W.~H. Guss, A.~Nichol, A.~Paino, N.~Tezak, J.~Tang, I.~Babuschkin, S.~Balaji,
  S.~Jain, W.~Saunders, C.~Hesse, A.~N. Carr, J.~Leike, J.~Achiam, V.~Misra,
  E.~Morikawa, A.~Radford, M.~Knight, M.~Brundage, M.~Murati, K.~Mayer,
  P.~Welinder, B.~McGrew, D.~Amodei, S.~McCandlish, I.~Sutskever, and
  W.~Zaremba, ``Evaluating large language models trained on code,'' 2021.

\bibitem{touvron2023llama}
H.~Touvron \emph{et~al.}, ``Llama 2: Open foundation and fine-tuned chat
  models,'' 2023.

\bibitem{luo2023wizardcoder}
Z.~Luo, C.~Xu, P.~Zhao, Q.~Sun, X.~Geng, W.~Hu, C.~Tao, J.~Ma, Q.~Lin, and
  D.~Jiang, ``Wizardcoder: Empowering code large language models with
  evol-instruct,'' \emph{arXiv preprint arXiv:2306.08568}, 2023.

\bibitem{phind-codellama-34b-v2}
\BIBentryALTinterwordspacing
Phind. (2023) Phind-codellama-34b-v2. [Online]. Available:
  \url{https://huggingface.co/Phind/Phind-CodeLlama-34B-v2}
\BIBentrySTDinterwordspacing

\bibitem{Barke2022GroundedCH}
S.~Barke, M.~B. James, and N.~Polikarpova, ``Grounded copilot: How programmers
  interact with code-generating models,'' \emph{ArXiv}, vol. abs/2206.15000,
  2022.

\bibitem{Dderlein2022PilotingCA}
J.-B. D{\"o}derlein, M.~Acher, D.~E. Khelladi, and B.~Combemale, ``Piloting
  copilot and codex: Hot temperature, cold prompts, or black magic?''
  \emph{ArXiv}, vol. abs/2210.14699, 2022.

\bibitem{Sarkar2022WhatII}
A.~Sarkar, A.~D. Gordon, C.~Negreanu, C.~Poelitz, S.~S. Ragavan, and B.~G.
  Zorn, ``What is it like to program with artificial intelligence?''
  \emph{ArXiv}, vol. abs/2208.06213, 2022.

\bibitem{guo2023longcoder}
D.~Guo, C.~Xu, N.~Duan, J.~Yin, and J.~McAuley, ``Longcoder: A long-range
  pre-trained language model for code completion,'' in \emph{International
  Conference on Machine Learning}.\hskip 1em plus 0.5em minus 0.4em\relax PMLR,
  2023, pp. 12\,098--12\,107.

\bibitem{white2023chatgpt}
J.~White, S.~Hays, Q.~Fu, J.~Spencer-Smith, and D.~C. Schmidt, ``Chatgpt prompt
  patterns for improving code quality, refactoring, requirements elicitation,
  and software design,'' \emph{arXiv preprint arXiv:2303.07839}, 2023.

\bibitem{chen2023teaching}
X.~Chen, M.~Lin, N.~Sch{\"a}rli, and D.~Zhou, ``Teaching large language models
  to self-debug,'' \emph{arXiv preprint arXiv:2304.05128}, 2023.

\bibitem{yasunaga2021break}
M.~Yasunaga and P.~Liang, ``Break-it-fix-it: Unsupervised learning for program
  repair,'' in \emph{International conference on machine learning}.\hskip 1em
  plus 0.5em minus 0.4em\relax PMLR, 2021, pp. 11\,941--11\,952.

\bibitem{zhang2022repairing}
J.~Zhang, J.~Cambronero, S.~Gulwani, V.~Le, R.~Piskac, G.~Soares, and
  G.~Verbruggen, ``Repairing bugs in python assignments using large language
  models,'' \emph{arXiv preprint arXiv:2209.14876}, 2022.

\bibitem{sobania2023analysis}
D.~Sobania, M.~Briesch, C.~Hanna, and J.~Petke, ``An analysis of the automatic
  bug fixing performance of chatgpt,'' in \emph{2023 IEEE/ACM International
  Workshop on Automated Program Repair (APR)}.\hskip 1em plus 0.5em minus
  0.4em\relax IEEE, 2023, pp. 23--30.

\bibitem{khan2022automatic}
J.~Y. Khan and G.~Uddin, ``Automatic code documentation generation using
  gpt-3,'' in \emph{Proceedings of the 37th IEEE/ACM International Conference
  on Automated Software Engineering}, 2022, pp. 1--6.

\bibitem{schafer2023empirical}
M.~Sch{\"a}fer, S.~Nadi, A.~Eghbali, and F.~Tip, ``An empirical evaluation of
  using large language models for automated unit test generation,'' \emph{IEEE
  Transactions on Software Engineering}, 2023.

\bibitem{chen2022codet}
B.~Chen, F.~Zhang, A.~Nguyen, D.~Zan, Z.~Lin, J.-G. Lou, and W.~Chen, ``Codet:
  Code generation with generated tests,'' \emph{arXiv preprint
  arXiv:2207.10397}, 2022.

\bibitem{chen2023lm4hpc}
L.~Chen, P.-H. Lin, T.~Vanderbruggen, C.~Liao, M.~Emani, and B.~de~Supinski,
  ``Lm4hpc: Towards effective language model application in high-performance
  computing,'' in \emph{OpenMP: Advanced Task-Based, Device and Compiler
  Programming}, S.~McIntosh-Smith, M.~Klemm, B.~R. de~Supinski, T.~Deakin, and
  J.~Klinkenberg, Eds.\hskip 1em plus 0.5em minus 0.4em\relax Cham: Springer
  Nature Switzerland, 2023, pp. 18--33.

\bibitem{ding2023hpc}
X.~Ding, L.~Chen, M.~Emani, C.~Liao, P.-H. Lin, T.~Vanderbruggen, Z.~Xie,
  A.~Cerpa, and W.~Du, ``Hpc-gpt: Integrating large language model for
  high-performance computing,'' in \emph{Proceedings of the SC'23 Workshops of
  The International Conference on High Performance Computing, Network, Storage,
  and Analysis}, 2023, pp. 951--960.

\bibitem{chen2024ompgpt}
L.~Chen, A.~Bhattacharjee, N.~Ahmed, N.~Hasabnis, G.~Oren, V.~Vo, and
  A.~Jannesari, ``Ompgpt: A generative pre-trained transformer model for
  openmp,'' \emph{arXiv preprint arXiv:2401.16445}, 2024.

\bibitem{chen2023data}
L.~Chen, X.~Ding, M.~Emani, T.~Vanderbruggen, P.~hung Lin, and C.~Liao, ``Data
  race detection using large language models,'' 2023.

\bibitem{mankowitz2023faster}
D.~J. Mankowitz, A.~Michi, A.~Zhernov, M.~Gelmi, M.~Selvi, C.~Paduraru,
  E.~Leurent, S.~Iqbal, J.-B. Lespiau, A.~Ahern \emph{et~al.}, ``Faster sorting
  algorithms discovered using deep reinforcement learning,'' \emph{Nature},
  vol. 618, no. 7964, pp. 257--263, 2023.

\end{thebibliography}

\end{document}